\documentclass[doublecol]{epl2}

\title{Impurity and edge roughness scattering in armchair graphene nanoribbons: Boltzmann approach}
\shorttitle{Title} 

\author{Hengyi Xu\and Thomas Heinzel}
\shortauthor{F. Author \etal}

\institute{
   Condensed Matter Physics Laboratory, Heinrich-Heine-Universit\"at,
Universit\"atsstr.1, 40225 D\"usseldorf, Germany\\
}

\abstract{The conductivity of armchair graphene nanoribbons in the presence of short-range impurities and edge roughness is studied theoretically using the Boltzmann transport equation for quasi-one-dimensional systems. As the number of occupied subbands increases, the conductivity due to short-range impurities converges towards the two-dimensional case.  Calculations of the magnetoconductivity confirm the edge-roughness-induced dips at cyclotron radii close to the ribbon width suggested by the recent quantum simulations.  }

\begin{document}

\maketitle

\section{Introduction}
The theoretical descriptions of electron transport in graphene sheets with various scattering sources, such as charged impurities in
substrates, microscopic corrugations or short-range resonant scatters, have been frequently  based on the Boltzmann approach.
Such studies, in particular the electron density dependence of the conductivity, are of fundamental interest since they help identifying the dominant scattering sources. \cite{Peres2011,Sarma2011} It has been verified systematically that the Boltzmann approach works quite well to describe the transport in broad parameter ranges for both single and  double layer graphene. \cite{Xu2011PRB} As the width of the graphene strips is decreased, graphene nanoribbons (GNRs) are fromed, where size-dependent effects, for example the inhomogeneous electron density \cite{Xu2010PRB} or the edge roughness, become relevant for the transport properties. In this case, the transport has so far been described within the framework of the Landauer-B\"{u}ttiker model with the aid of Green's function techniques. \cite{Lewenkopf2008,Xu2008PRB} It is well established that in GNRs, edge disorder can contribute significantly to the scattering \cite{Areshkin2007,Evaldsson2008,Xu2009PRB,Han2007} which in wide structures is governed by a combination of scattering at charged impurities and resonant scattering at short-range defects. Edge disorder has been suggested as the source of the transport gap in narrow GNRs around the charge neutrality point.\cite{Han2007,Stampfer2009,Liu2009,Todd2009} Furthermore, a typical size effect, the so-called edge-roughness-induced magnetoconductance dip (ERID) in GNRs has been studied by numerical quantum simulations, which is interpreted as a magnetic-field-enhanced diffusive scattering when the electron trajectory grazes at the edges. \cite{Xu2012EPL} On the other hand, the Boltzmann approach has been applied to treat a variety of scattering sources in conventional quasi-one-dimensional (Q1D) systems, for example quantum wires. \cite{Akera1991,Bruus1993,Feilhauer2011}. However, only a few aspects of  transport in GNRs have so far been studied within the Boltzmann model. \cite{Huang2010JAP,Huang2011PRB}

In the present paper, we apply the linear Boltzmann equation to armchair GNRs and determine its transport properties in the presence of $\delta$-type short-range impurities and edge roughness. The magnetoconductivity in wide GNRs with rough edge roughness is studied as well.

\section{Model and theory}
We start with the Dirac Hamiltonian
\begin{equation}
H=\hbar v_F(\sigma_x\tau_zk_x+\sigma_yk_y)
\end{equation}
with Fermi velocity $v_F\approx 10^6m/s$ and Pauli matrices $\sigma_{x,y}$ and $\tau_z$ acting on the $A/B$ sublattice and $K/K'$ valley spaces, respectively. The energy spectrum of GNRs depends on the nature of their edges, namely zigzag or armchair. Within the present work, we restrict ourselves to metallic armchair GNRs (AGNRs). For this system, the boundary conditions imposed on the wave function, namely $\Psi_A(x=0)=\Psi_B(x=0)=\Psi_A(x=W)=\Psi_B(x=W)=0$, give rise to the allowed transverse wave vectors as
\begin{equation}
k_n=\frac{n\pi}{W}-\frac{4\pi}{3a} \label{allkn}
\end{equation}
with $a=0.246 \mathrm{nm}$ being the lattice constant of graphene. $W$ is the width of AGNRs. The integer $n$ is of the order of $W/a$ for the energetically lowest modes. Throughout this text, we denote the energy $\epsilon$ normalized to $\hbar v_F$ as $\widetilde{\epsilon}\equiv \epsilon/(\hbar v_F)$ with $\widetilde{\epsilon}^2=k_n^2+k^2$. The normalized wave function for the $n$th subband reads \cite{CastroNeto2009,WurmNJP2009}
\begin{equation}
\Psi(\mathbf{r})=\frac{e^{iky}}{\sqrt{4WL}}\left(\begin{array}{c}
e^{ik_nx}\\
 \frac{k_n+ik}{\widetilde{\epsilon}_{nk}}e^{ik_nx}\ \\
 -e^{-ik_nx}\\\
-\frac{k_n+ik}{\widetilde{\epsilon}_{nk}}e^{-ik_nx}\
\end{array} \right)\label{wavefunc}
\end{equation}
which is a mixture of two Dirac points $\mathbf{K}=(4\pi/(3a),0)=(K,0)$ and $\mathbf{K'}=(-4\pi/(3a),0)=(-K,0)$.
Here we choose the wave vectors in the $x$-direction to be quantized and the transport is oriented along $y$-direction. $L$ is the length of the system.

To describe the transport properties of GNRs, we adopt the linearized Boltzmann equation describing the general Q1D system
\begin{equation}
-\frac{eE_y}{\hbar}\frac{\partial f_{nk}^0(\epsilon_{nk})}{\partial k}=\sum_{n'}\sum_{k'}\mathcal{W}_{n'k'nk}\left[ f_{n'k'}-f_{nk}\right]\label{boltz}
\end{equation}
where $E_y$ is the applied electric field along the transport direction, $f_{nk}$ is the distribution function of a state with wave vector $k$ and energy $\epsilon_{nk}$ in the $n$th subband, and the superscript ``0'' denotes the equilibrium distribution. According to Fermi's Golden Rule, the scattering probability due to the perturbation potential is given by
\begin{equation}
\mathcal{W}_{n'k'nk}=\frac{2\pi}{\hbar}|\langle n',k'| U|n,k\rangle|^2\delta(\epsilon_{n'k'}-\epsilon_{nk})\label{femirule}
\end{equation}
Using the relaxation time approximation, the nonequilibrium distribution function can be written as
\begin{equation}
f_{nk}(\epsilon_{nk})=f_{nk}^0(\epsilon_{nk})-eE_xv_n(\epsilon_{nk})\tau_n(\epsilon_{nk})\delta(\epsilon_{nk}-E_F)\label{ffn}
\end{equation}
with Fermi energy $E_F$ and the relaxation time $\tau_n$ for the state in the $n$th subband. The velocity for the $n$th subband $v_n=(1/\hbar){\partial \epsilon_{nk}}/{\partial k}=v_F{k}/{\sqrt{(k_n)^2+k^2}}$ for the linear spectrum of graphene.
Inserting Eq. (\ref{ffn}) into Eq. (\ref{boltz}), the Boltzmann equation at zero temperature can be written as
\begin{eqnarray}
\frac{k}{\widetilde{\epsilon}_{nk}}\delta(\epsilon_{nk}-E_F)&=&\sum_{n',k'}\mathcal{W}_{n'k'nk}\left[\frac{k }{\widetilde{\epsilon}_{nk}}\tau_n(\epsilon_{nk})\delta(\epsilon_{nk}-E_F)\right.\nonumber\\
&&\left.- \frac{k'}{\widetilde{\epsilon}_{n'k'}}\tau_{n'}(\epsilon_{n'k'})\delta(\epsilon_{n'k'}-E_F)\right].\label{boltz1}
\end{eqnarray}
Multiplying both sides of Eq. (\ref{boltz1}) by $k$ and summing over $k$, we obtain after some algebra
\begin{equation}\label{}
    k_F^n=\frac{2\pi}{\hbar}\sum_{n'}\mathcal{T}_{nn'}\tau_{n'}(E_F)
\end{equation}
where $k_F^n$ is Fermi wave vector in the $n$th subband. The transition matrix element $\mathcal{T}_{nn'}$ is defined as
\begin{eqnarray}
  \mathcal{T}_{nn'}&=&\frac{\pi\hbar v_F}{L}\sum_{k'}\sum_{k}\left[ \delta_{nn'} \sum_{\mu} |\langle \mu,k'| U|n,k\rangle|^2\right.\nonumber\\
  &\times&\frac{k^2}{\widetilde{\epsilon}_{nk}}\delta(\epsilon_{nk}-E_{F})\delta(E_{\mu k'}-E_{F}) \nonumber\\
&-&\left. |\langle n',k'| U|n,k\rangle|^2\frac{kk'}{\widetilde{\epsilon}_{n'k'}}\delta(\epsilon_{nk}-E_{F})\delta(\epsilon_{n' k'}-E_{F})\right]\nonumber\\
\end{eqnarray}
with the summation running over the mode index $\mu$. The Boltzmann conductivity for GNRs is then given by
\begin{equation}\label{}
    \sigma(E_F)=\frac{2e^2}{h}\frac{\hbar^2 v_F^2}{\pi E_F}\frac{1}{W}\sum_{n,n'} k_F^nk_F^{n'}(\mathcal{T}^{-1})_{nn'}\label{sigma1}
\end{equation}
For nonzero temperature, the conductivity is obtained from
\begin{equation}\label{}
    \sigma=\int d\epsilon  \left(-\frac{\partial f(\epsilon)}{\partial \epsilon}\right)\sigma(\epsilon)\label{sigma2}
\end{equation}

We proceed by describing how the scattering potentials have been implemented in this formalism. First, we consider $\delta$-type impurities in the form of
\begin{equation}\label{}
    U=\gamma\sum_{j=1}^{N_I}\delta(x-x_j)\delta(y-y_j)
\end{equation}
where $\gamma$ and $N_I$ are the strength and the number of impurities, respectively. Thus the matrix element squared of the perturbation
is evaluated as
\begin{eqnarray}
&&|\langle n'k'|U|nk\rangle|^2=\frac{\gamma^2}{4W^2L^2}\nonumber\\
&\times&\sum_{j=1}^{N_j}\cos^2\frac{(n-n')\pi x_j}{W}\left|1+\frac{(k_{n'}-ik')(k_n+ik)}{\widetilde{\epsilon}_{n'k'}\widetilde{\epsilon}_{nk}} \right|^2\nonumber\\
&=&\frac{\gamma^2n_i}{4WL}(1+\delta_{nn'})\left(1+\frac{k_{n'}k_n+k'k}{\widetilde{\epsilon}_{n'k'}\widetilde{\epsilon}_{nk}}\right)
\end{eqnarray}
with $n_i=N_I/WL$. The transition matrix elements are finally given by
\begin{eqnarray}
\mathcal{T}_{nn'}&=&\frac{\gamma^2n_i}{4\pi W}\delta_{nn'}\left(\sum_\mu \left[(1+\delta_{n\mu})\left(\frac{E_F}{\hbar^2 v_F^2}+\frac{k_\mu k_n}{E_{F}}\right)\frac{k_F^{n}}{k_F^\mu}
\right]\right.\nonumber\\
&&\left.-\frac{k_F^{n'}k_F^n}{{E}_{F}} \right)-\frac{\gamma^2n_i}{4\pi W}\frac{k_F^{n'}k_F^n}{{E}_{F}}.\label{knn1}
\end{eqnarray}
Using  Eq. (\ref{knn1}) and Eq. (\ref{sigma1}) we obtain for the conductivity of GNRs the expression
\begin{eqnarray}
\sigma&=&\frac{8e^2}{h}\frac{(\hbar v_F)^2}{\gamma^2n_i}\sum_{n,n'}\left\{\left[\sum_\mu\left((1+\delta_{n\mu})(\widetilde{E}^2_F+k_\mu k_n)
\frac{k_F^n}{k_F^\mu}\right)\right.\right.\nonumber\\&&\left.\left.-k_F^{n'}k_F^n\right]\delta_{nn'}-k_F^{n'}k_F^n\right\}^{-1}_{nn'}k_F^nk_F^{n'}\label{sigma_delta}
\end{eqnarray}
For a large number of subbands, i.e., $\mathcal{N}\gg 1$, we have $k_\mu,k_n\ll \widetilde{E}_F$ and $k_F^n,k_F^{n'}\sim \widetilde{E}_F$. In this approximation, Eq. (\ref{sigma_delta}) converges to the well-known result for the extended case
\begin{equation}
\sigma=\frac{8e^2}{h}\frac{\hbar^2 v_F^2}{\gamma^2n_i} \label{sigma_delta2}
\end{equation}
which is independent of the carrier concentration. \cite{Sarma2011,Xu2011PRB}

As a second scattering mechanism, we study the effects of edge roughness in the absence of magnetic fields. The edge roughness is parametrized by
$\Delta(y){\partial V(x)}/{\partial x}$, an expression which has been applied before to model rough semiconductor quantum wires and interfaces. \cite{Ferry1985PRB,Ferrybook} $V(x)$ is the one-dimensional confinement potential which can be modeled by a finite mass term in the Dirac Hamiltonian. \cite{Peres2009JPC}  $\Delta(y)$ is a function describing the potential fluctuation of GNRs and characterized by $\langle \Delta(y) \rangle=0$ and the autocovariance function $\langle \Delta(y)\Delta(y')\rangle=\Delta^2\exp[-(y-y')^2/\Lambda^2]$ with $\Lambda$ being the correlation length. Furthermore, $\langle\cdots\rangle$ denotes position averaging.

To evaluate the perturbation matrix element, we define the function $\Xi$ related to the $x$-components of the wave functions as
\begin{equation}
 \Xi_{n'n}=\frac{1}{W}\int_{0(W)}^{-\infty(+\infty)} dx \phi_{n'}^*(x) \frac{\partial V(x)}{\partial x}\phi_n(x)
\end{equation}
where $\phi_n(x)$ denotes one of the components of wave function in Eq. (\ref{wavefunc}). For the hard-wall confinement potential present in GNRs,
this function can be  expressed as
\begin{equation}
\Xi_{n'n}=\left.-\frac{1}{W}\frac{\hbar v_F}{2\widetilde{E}_F}\left[\frac{\partial \phi^*_{n'}}{\partial x}\frac{\partial \phi_{n}}{\partial x}\right]\right|_{x=0,W}\label{eq18}
\end{equation}
It is noteworthy that a linear form for the matrix elements of the edge roughness perturbation has been used, which however neglects the interband scattering. \cite{Fang2008PRB,Huang2011PRB}
Using Eq. (\ref{eq18}), the square of the matrix element for edge roughness reads
\begin{eqnarray}
|\langle n',k'|U|n,k\rangle|^2&=&\frac{\pi^{9/2}n'^2n^2}{8W^6}\frac{(\hbar v_F)^2}{\widetilde{E}^2_F}\left(1+\frac{k_nk_{n'}+kk'}{\widetilde{\epsilon}_{nk}\widetilde{\epsilon}_{n'k'}} \right)\nonumber\\
&&\times\frac{\Lambda\Delta^2}{L} \exp[-\Lambda^2(k-k')^2/4]
\end{eqnarray}
where we have used the Gaussian integral in the evaluation of part in the $y$-direction. In the case of small correlation length $\Lambda\ll \lambda_F$, the transition matrix element has the form
\begin{eqnarray}
\mathcal{T}_{nn'}&=&\frac{\pi^{7/2}\Lambda\Delta^2}{8W^6}\frac{\hbar v_F}{\widetilde{E}_F}n^2\left[\sum_\mu \mu^2\left( 1+\frac{k_nk_\mu}{\widetilde{E}^2_F}\right)\frac{k_F^n}{ k_F^\mu}\delta_{nn'}\right.\nonumber\\
&&\left.-n'^2\frac{k_F^{n'}k_F^n}{\widetilde{E}^2_F}\right]
\end{eqnarray}
This results in a conductivity given by
\begin{eqnarray}
\sigma&=&\frac{16e^2}{h}\frac{W^5}{\pi^{9/2}\Lambda\Delta^2}\sum_{n,n'}k_F^nk_F^{n'}\nonumber\\
&\times&\left[n^2\sum_\mu \mu^2\left( 1+\frac{k_nk_\mu}{\widetilde{E}^2_F}\right)\frac{k_F^n}{ k_F^\mu}\delta_{nn'}-n^2n'^2\frac{k_F^{n'}k_F^n}{\widetilde{E}^2_F}\right]^{-1}_{nn'}.\nonumber\\ \label{sigma_edge}
\end{eqnarray}
Regarding the diagonal contributions of the transition rate matrix, it is convenient to write the inverse of the relaxation time for the $n$th subband as
\begin{equation}
\frac{1}{\tau_n}=\frac{\pi^{9/2}}{ 4W^6}\frac{\hbar v_F^2}{E_F}\Lambda\Delta^2n^2\sum_{\mu}\mu^2\left[\frac{1}{k_F^\mu}+\frac{k_\mu k_n/k_F^\mu}{\widetilde{E}^2_F} \right].\label{relaxt1}
\end{equation}
For a large number of occupied subbands ($\mathcal{N}\gg 1$), the second term in the bracket can be neglected since $k_{n,\mu}\ll \widetilde{E}_F$, and the summation can be replaced by an integral. Eq.(\ref{relaxt1}) can then be written as
\begin{equation}
\frac{1}{\tau_n}\approx\frac{\pi^{5/2}}{16W^3}\Lambda\Delta^2v_F\widetilde{E}_Fn^2\label{relaxt2}
\end{equation}
which shows a striking similarity for the corresponding results reported for semiconductor quantum wires. \cite{Akera1991}. Furthermore, this relaxation time results in a conductivity in the limit of $\mathcal{N}\gg 1$ of
\begin{equation}
\sigma\approx\frac{32e^2}{h}\frac{ 1}{3\pi^{1/2}}\frac{W^2}{\Lambda\Delta^2\widetilde{E}_F}.\label{sigma_edge2}
\end{equation}

Up to now, we have looked at the transport with edge roughness in the absence of magnetic fields $B$ and now continue by including it and discussing its semiclassical effects. A weak magnetic field may homogenize the contributions of the occupied subbands to the overall conductivity and result in a reduction of the magnetoconductivity when the cyclotron radius $r_c\sim W$. This effect has been verified numerically by recent quantum calculations in GNRs \cite{Xu2012EPL} but, to the best of our knowledge, has so far not yet been observed. Here the maximum reduction of the relaxation time $\tau$ in magnetic fields can be roughly estimated by averaging over all occupied modes, resulting in
\begin{equation}
\frac{1}{\tau(B>0)}\approx\frac{\pi^{1/2}}{48 W}\Lambda\Delta^2v_F\widetilde{E}^3_F.
\end{equation}
The magnetoconductivity near the dip is then given by
\begin{equation}
\sigma(B>0)\approx\frac{48e^2}{h}\frac{1}{\pi^{1/2}}\frac{W}{\Lambda\Delta^2\widetilde{E}_F^2},\label{sigma_edge_B}
\end{equation}
which holds for $\mathcal{N}\gg 1$.

\section{Numerical results and discussion}
\begin{figure}[tbp]
\includegraphics[width=75mm]{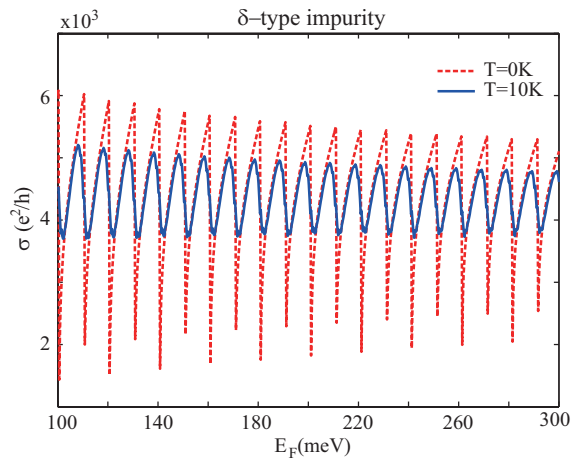}
\caption{(Colour online) The numerical conductivity of $\delta$-type impurities in armchair GNRs with $W=180$nm plotted versus the Fermi energy for two different temperatures T. We have chosen the parameters as $\overline{\gamma}=0.9$ and $\overline{n}_i=0.5$, where $\overline{\gamma}=\gamma (\hbar v_F)^2/E_F$ and $\overline{n}_i=n_i\lambda_F^2$.}
\label{Fig1}
\end{figure}

In Fig. \ref{Fig1} we show the conductivity for $\delta$-type impurities according to Eq.(\ref{sigma_delta}) as a function of Fermi energy. For different degrees of disorder, the conductivity shows very similar features while the amplitude of the conductivity depends on the disorder parameters. Prominent quantum oscillations at zero temperature are observed, i.e., the conductivity drops rapidly as a new scattering channel is opened and increases again until the Fermi energy hits the next subband at larger energies. The oscillations are smeared by finite temperature to some extent. In the whole range of Fermi energies studied, the average conductivity remains independent of the carrier concentration, which is consistent with the two-dimensional case. \cite{Sarma2011,Xu2011PRB}

\begin{figure}[tbp]
\includegraphics[width=75mm]{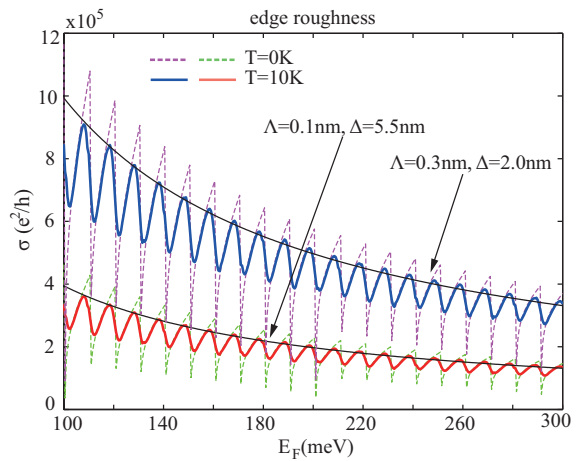}
\caption{(Colour online) The Boltzmann conductivity as a function of the Fermi energy for different edge roughness in AGNRs. The dashed and solid lines correspond to zero temperature and a temperature of $T=10 \mathrm{K}$, respectively. The smooth solid lines are calculated from Eq.(\ref{sigma_edge2}) in the limit of $\mathcal{N}\gg 1$. }
\label{Fig2}
\end{figure}

Fig. \ref{Fig2} shows the conductivity for edge roughness as a function of Fermi energy calculated from Eq. (\ref{sigma_edge}). The parameter values chosen for $\Lambda$ and $\Delta$ correspond to short-range defects, for instance, a few atoms missing at the GNR edges, as widely assumed in simulations of edge disorder \cite{Evaldsson2008,Mucciolo2009,Xu2009PRB,Xu2012EPL}. The correlation length ensures that $\Lambda\ll \lambda_F$ over the whole range of Fermi energies. The Boltzmann conductivity at nonzero temperature (indicated by the solid lines) shows suppressed quantum fluctuations in comparison with the zero-temperature cases (indicated by the dashed line). The overall conductivity decreases as the Fermi energy increases. Since the correlation length $\Lambda$ and edge position fluctuation amplitude $\Delta$ increase relative to the Fermi wavelength as $E_F$ is increased, this behavior is similar to that one found in conventional quantum wires \cite{Akera1991,Feilhauer2011}. In the case of large number of subbands, $\mathcal{N}\gg 1$, the results from Eq. (\ref{sigma_edge2}) (indicated by solid lines) exhibit the same overall trends and agree well with the exact ones except the absence of the quantum oscillations.

The Fermi energies in Fig. \ref{Fig1} and \ref{Fig2} correspond to numbers of subbands between $10$ and $30$. For smaller Fermi energies, i.e. a few occupied modes, conductivity shows more prominent quantum fluctuations and may deviate considerably from the asymptotic expressions. Moreover, it should be noted that our calculations based on the Boltzmann approach is valid for the case where the interband scattering is rather strong. This is the case when the mean free path is considerably shorter than the length of the graphene ribbon.

\begin{figure}[tbp]
\includegraphics[width=75mm]{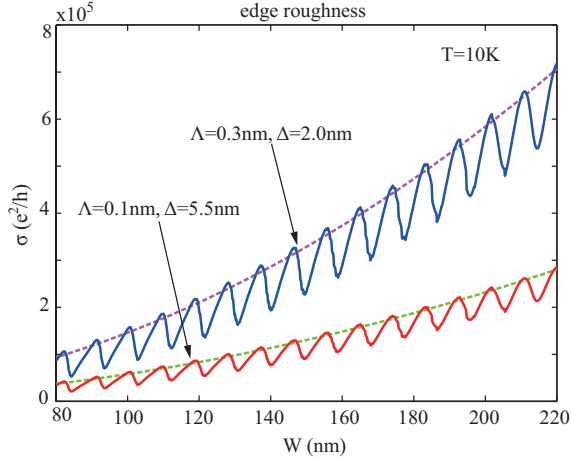}
\caption{(Colour online) The width dependence of the conductivity for different edge roughness in AGNRs with $E_F=200\mathrm{meV}$ and temperature $T=10\mathrm{K}$. The solid lines show the exact conductivity from Eq.(\ref{sigma2}) by the $\mathcal{T}$ matrix inversion, and dashed lines correspond to the limit $\mathcal{N}\gg 1$ from Eq.(\ref{sigma_edge2}).}
\label{Fig3}
\end{figure}

Fig. \ref{Fig3} shows the Boltzmann conductivity as a function of the GNR width for different edge roughness parameters. Here, only results for $T>0$ are presented. The overall conductivity for two roughness levels exhibits a parabolic dependence on the width, superimposed by quantum oscillations. This quadratic behavior may be seen more clearly from the analytical expression Eq. (\ref{sigma_edge2}), as illustrated by the dashed lines.

\begin{figure}[tbp]
\includegraphics[width=75mm]{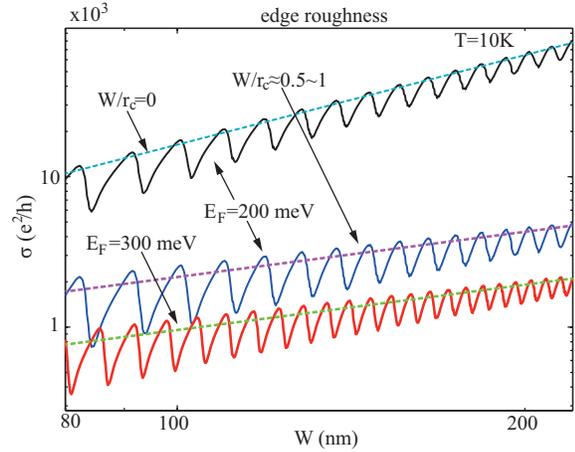}
\caption{(Colour online) The magnetoconductivity around ERID as a function of the width for different Fermi energies at finite temperature $T=10\mathrm{K}$. (Note the logarithmic scale.) The zero-field conductivity with $E_F=200 \mathrm{meV}$  is also plotted at the top. The solid and dashed lines correspond to the results from Eq.(\ref{sigma_edge}) and Eq.(\ref{sigma_edge_B}) for $\mathcal{N}\gg 1$, respectively. $r_c$ is the cyclotron radius.}
\label{Fig4}
\end{figure}

In the following, we give a rough estimate of the GNR conductivity in magnetic fields with amplitudes close to the position of the ERID, i.e., $r_c\approx W$. A more exact calculation would have to rely upon a calculation of  the wave functions in magnetic fields, which can be obtained by solving the eigenequation of the Dirac Hamiltonian with magnetic fields included. \cite{Brey2006PRB,Alessandro2011PRB} This, however, should have only a marginal effect and we limit ourselves to the qualitative properties of the system close to the ERID.

In Fig. \ref{Fig4}, the conductivity at $T=10 \mathrm{K}$ is shown in the vicinity of the ERID for different Fermi energies. The parameters for edge roughness are fixed to $\Lambda=0.3 \mathrm{nm}$ and $\Delta=6 \mathrm{nm}$. For comparison, the corresponding zero-field conductivity with $E_F=200 \mathrm{meV}$ is plotted as well. The conductivity around the ERID increases linearly with the GNR width, in contrast to the parabolic dependence in the absence of a magnetic field. This linear relationship can be easily identified from Eq. (\ref{sigma_edge_B}) and is also illustrated by the dashed lines in Fig. \ref{Fig4}.  As a consequence of the distinctly different dependencies of $\sigma$ on  W, the ERID can be expected to be more pronounced in wider GNRs. This feature is in qualitative agreement with our previous quantum simulations. \cite{Xu2012EPL}.

We conclude this analysis by commenting on the observability of the ERID in realistic GNRs. First of all, the length of the GNR is irrelevant in the present treatment since diffusive transport has been assumed. Second, we have restricted ourselves to the case of rather small correlation lengths for the edge roughness. It is self-evident that a large correlation length suppresses the ERID in view the reduced diffusiveness of the scattering at the edges. Furthermore, the bulk disorder must remain at a sufficiently low level as indicated by the quantum simulations before, such that it does not mask completely the edge roughness scattering.
We moreover expect that qualitatively, the ERID does not depend much in the the type of edges, even though numerical simulations suggest that zigzag GNRs are more robust with respect to edge disorder. \cite{Xu2012EPL} Similar analytical expressions for zigzag GNRs are possible in principle but more complicated due to the presence of surface states and the interdependence of the transverse and longitudinal wave vectors.

In summary, we have studied the transport properties of AGNRs with short-range impurities and edge roughness within the framework given by the Boltzmann equation. An edge-roughness-induced magnetoconductivity minimum suggested by the recent quantum calculations is confirmed by the Boltzmann results and should become observable experimentally if the correlation length of the edge roughness is not much larger than the Fermi wavelength and the bulk disorder is sufficiently low. It has been shown that the ERID induced by the magnetic-field-enhanced diffusive scattering at rough edges shows a behavior very similar to that one found in conventional semiconductor quantum wires, despite the fundamentally different energy dispersion.

\acknowledgments
H.X. and T.H. acknowledge financial support from Heinrich-Heine-Universit\"{a}t D\"{u}sseldorf.


\begin{thebibliography}{10}
\expandafter\ifx\csname url\endcsname\relax\def\url#1{\texttt{#1}}\fi

\bibitem{Peres2011}
\Name{Peres N. M.~R.} \REVIEW{Rev. Mod. Phys.}{82}{2011}{2673}.

\bibitem{Sarma2011}
\Name{Sarma S.~D., Adam S., Hwang E.~H. \and Rosso E.} \REVIEW{Rev. Mod.
  Phys.}{83}{2011}{407}.

\bibitem{Xu2011PRB}
\Name{Xu H., Heinzel T. \and Zozoulenko I.~V.} \REVIEW{Phys. Rev.
  B}{84}{2011}{115409}.

\bibitem{Xu2010PRB}
\Name{Xu H., Heinzel T., Shylau A.~A. \and Zozoulenko I.~V.} \REVIEW{Phys. Rev.
  B}{82}{2010}{115311}.

\bibitem{Lewenkopf2008}
\Name{Lewenkopf C.~H., Mucciolo E.~R. \and Castro~Neto A.~H.} \REVIEW{Phys.
  Rev. B}{77}{2008}{081410}.

\bibitem{Xu2008PRB}
\Name{Xu H., Heinzel T., Evaldsson M. \and Zozoulenko I.~V.} \REVIEW{Phys. Rev.
  B}{77}{2008}{245401}.

\bibitem{Areshkin2007}
\Name{Areshkin D.~A., Gunlycke D. \and White C.~T.} \REVIEW{Nano
  Lett.}{7}{2007}{204}.

\bibitem{Evaldsson2008}
\Name{Evaldsson M., Ihnatsenka S. \and Zozoulenko I.~V.} \REVIEW{Phys. Rev.
  B}{77}{2008}{165306}.

\bibitem{Xu2009PRB}
\Name{Xu H., Heinzel T. \and Zozoulenko I.~V.} \REVIEW{Phys. Rev.
  B}{80}{2009}{045308}.

\bibitem{Han2007}
\Name{Han M.~Y., Ozyilmaz B., Zhang Y. \and Kim P.} \REVIEW{Phys. Rev.
  Lett.}{98}{2007}{206805}.

\bibitem{Stampfer2009}
\Name{Stampfer C., Guettinger J., Hellmueller S., Molitor F., Ensslin K. \and
  Ihn T.} \REVIEW{Phys. Rev. Lett.}{102}{2009}{056403}.

\bibitem{Liu2009}
\Name{Liu X.~L., Oostinga J.~B., Morpurgo A.~F. \and Vandersypen L. M.~K.}
  \REVIEW{Phys. Rev. B}{80}{2009}{121407}.

\bibitem{Todd2009}
\Name{Todd K., Chou H.~T., Amasha S. \and Goldhaber-Gordon D.} \REVIEW{Nano
  Lett.}{9}{2009}{416}.

\bibitem{Xu2012EPL}
\Name{Xu H., Heinzel T. \and Zozoulenko I.~V.} \REVIEW{EPL (Europhysics
  Letters)}{97}{2012}{28008}.

\bibitem{Akera1991}
\Name{Akera H. \and Ando T.} \REVIEW{Phys. Rev. B}{11676}{1991}{43}.

\bibitem{Bruus1993}
\Name{Bruus H., Flensberg K. \and Smith H.} \REVIEW{Phys. Rev.
  B}{48}{1993}{11144}.

\bibitem{Feilhauer2011}
\Name{Feilhauer J. \and Mo\ifmmode~\check{s}\else \v{s}\fi{}ko M.}
  \REVIEW{Phys. Rev. B}{83}{2011}{245328}.

\bibitem{Huang2010JAP}
\Name{Huang D. \and Gumbs G.} \REVIEW{J. Appl. Phys.}{107}{2010}{103710}.

\bibitem{Huang2011PRB}
\Name{Huang D., Gumbs G. \and Roslyak O.} \REVIEW{Phys. Rev.
  B}{83}{2011}{115405}.

\bibitem{CastroNeto2009}
\Name{Neto A. H.~C., Guinea F., Peres N. M.~R., Novoselov K.~S. \and Geim
  A.~K.} \REVIEW{Rev. Mod. Phys.}{81}{2009}{109}.

\bibitem{WurmNJP2009}
\Name{Wurm J., Wimmer M., Adagideli ., Richter K. \and Baranger H.~U.}
  \REVIEW{New Journal of Physics}{11}{2009}{095022}.

\bibitem{Ferry1985PRB}
\Name{Goodnick S.~M., Ferry D.~K., Wilmsen C.~W., Liliental Z., Fathy D. \and
  Krivanek O.~L.} \REVIEW{Phys. Rev. B}{32}{1985}{8171}.

\bibitem{Ferrybook}
\Name{Ferry D.~K. \and Goodnick S.~M.} \Book{Transport in Nanostructures} 1st
  Edition (Cambridge University Press) 1997.

\bibitem{Peres2009JPC}
\Name{Peres N. M.~R., Rodrigues J. N.~B., Stauber T. \and dos Santos J. M.
  B.~L.} \REVIEW{Journal of Physics: Condensed Matter}{21}{2009}{344202}.

\bibitem{Fang2008PRB}
\Name{Fang T., Konar A., Xing H. \and Jena D.} \REVIEW{Phys. Rev.
  B}{78}{2008}{205403}.

\bibitem{Mucciolo2009}
\Name{Mucciolo E.~R., Castro~Neto A.~H. \and Lewenkopf C.~H.} \REVIEW{Phys.
  Rev. B}{79}{2009}{075407}.

\bibitem{Brey2006PRB}
\Name{Brey L. \and Fertig H.~A.} \REVIEW{Phys. Rev. B}{73}{2006}{195408}.

\bibitem{Alessandro2011PRB}
\Name{De~Martino A., H\"utten A. \and Egger R.} \REVIEW{Phys. Rev.
  B}{84}{2011}{155420}.

\end{thebibliography}
\end{document}